\documentclass[12pt]{iopart}

\usepackage{graphicx}

\usepackage{subeqn}
\usepackage{bbm}

\usepackage[svgnames]{xcolor}

\usepackage[T1]{fontenc}

\newcommand{\ket}[1]{\left|#1\right\rangle}

\newcommand{\ten}[1]{ \mathbbm{#1} }
\renewcommand{\vec}[1]{ \mathbf{#1} }
\newcommand{\strain}{u}
\renewcommand{\sup}{\uparrow}
\newcommand{\sdown}{\downarrow}

\newcommand{\eqref}[1]{\eref{#1}}
\begin{document}
\title{Strain-tuning of vacancy-induced magnetism in graphene nanoribbons}

\author{Daniel Midtvedt, Alexander Croy \footnote{email: croy@pks.mpg.de}}

\address{Max-Planck-Institut f\" ur Physik komplexer
Systeme, 01187 Dresden, Germany}

\begin{abstract}
    Vacancies in graphene lead to the appearance of localized electronic states with non-vanishing
    spin moments. Using a mean-field Hubbard model and an effective double-quantum dot description we investigate the
    influence of strain on localization and magnetic properties of the vacancy-induced states in
    semiconducting armchair nanoribbons. We find that the exchange splitting of a single vacancy
    and the singlet-triplet splitting for two vacancies can be widely tuned by applying uniaxial strain,
    which is crucial for spintronic applications.
\end{abstract}

% PACS number(s): 

\maketitle

%%%%%%%%%%%%%%%%%%%%%%%%%%%%%%%%%%%%%%%%%%%%%%%%%%%%%%%%%%%%%%%%%%%%%%%%%%%%%%%
%% INTRO
%%%%%%%%%%%%%%%%%%%%%%%%%%%%%%%%%%%%%%%%%%%%%%%%%%%%%%%%%%%%%%%%%%%%%%%%%%%%%%%
\section{Introduction}
Spintronics is an interesting prospective technology where spin is used as the information carrier (in contrast to electronic charge in the case of electronics) \cite{awfl07,fe08}. Graphene has been proposed as a promising material for realizing this technology \cite{haka+14}. A small spin-orbit coupling facilitates long spin decoherence times and makes information processing based on spin feasible\cite{trbu+07,drbu15b}. Magnetic order can can arise in graphene 
nanoflakes\cite{ez07,fepa07,wame+08,drbu15} and in connection with localized edge states in zig-zag nanoribbons\cite{fuwa+96}. Further, point defects in graphene, such as vacancies or chemisorpted hydrogen, have been proposed for realizations of spintronic devices\cite{yahe07,ya10}, since those defects support (quasi) localized electronic states and can induce magnetism \cite{wa02,pegu+06,pafe+08,nase+12}. In practice, such defects can be created by using electron or ion beams\cite{ya10,krba07}.

The magnetic order associated with the magnetic moments depends on the spatial configuration of the defects \cite{pafe+08,pimo+08}. For gapped graphene structures, such as nanoflakes \cite{drbu15} or armchair graphene nanoribbons (AGNRs) \cite{pafe+08}, the defect states are exponentially localized and well separated from the rest of the energy spectrum. Each defect can therefore be considered as a quantum dot (QD).

Two such localized non-degenerate defect states can be thought of as a basic unit of spin-based information processing. In \cite{drbu15}, it was shown that a system of two vacancies can be described as a double quantum dot (DQD) system. Further, it was shown that the exchange coupling in such systems can be tuned using a magnetic field.

Such \emph{in situ} tuning of the magnetic properties is essential for the realization of spintronic devices. In this work, we propose a complementary approach to achieve such tuning based on strain. The electronic properties of graphene are sensitive to strain\cite{peca09,peca+09,lugu10}, and it is known that GNRs have a strain-tunable band-gap\cite{cagu+09,lugu10}. However, there are only a few studies on the control of magnetism by strain, which so far focus on graphene quantum dots\cite{vipe+09,chyu+15}, adatoms\cite{huyu+11,shko+13} and magnetic impurities\cite{pogo+12,godu+13}. The influence of compressional strain on vacancy-induced magnetism in bulk graphene was studied in Ref.\ \cite{sari+12}.

We use AGNRs as a model system, and characterize the effect of strain on single and double vacancies in such systems. For a single vacancy we calculate the exchange splitting within a mean-field approach to the respective Hubbard model and show that it follows the strain-dependence of the band-gap. In the case of two vacancies, we use the localized defect states to setup a double quantum dot (DQD) model\cite{drbu15,lihe+14}. From this model
we calculate the singlet-triplet energy difference and find that it depends exponentially on the product of the distance of
vacancies and the applied strain. 
Our results suggest that strain can be used to non-invasively manipulate the properties of magnetic point defects in graphene. This is expected to be of crucial importance for spintronic applications. In particular, for applications consisting of arrays of such DQDs, strain provides a possibility to access and tune each DQD individually. Moreover, by
combining vacancy-induced magnetism with mechanical resonators, magneto-mechanic devices based on this principle are within reach.

\begin{figure}[b]
  \centering
  \includegraphics[width=0.49\columnwidth]
             {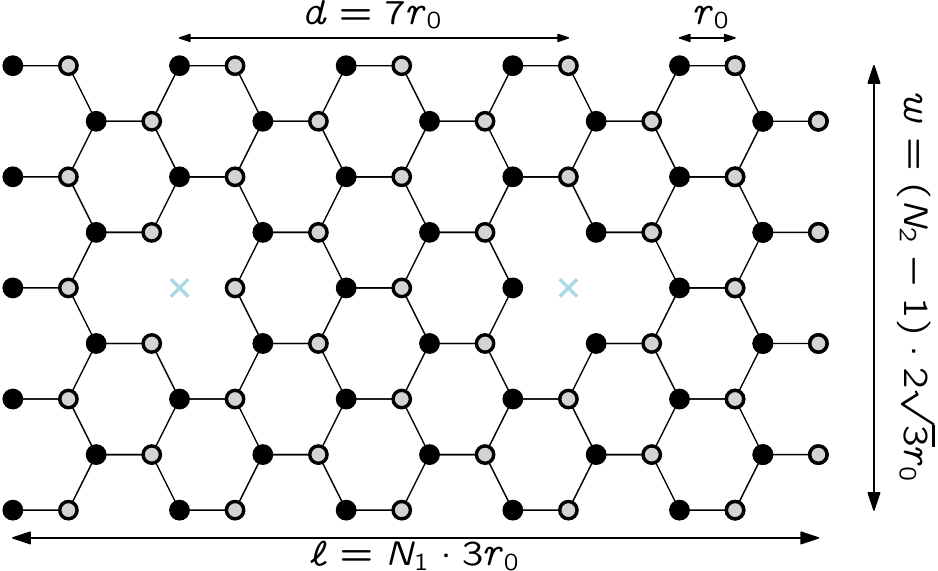}           
\caption{Structure of an armchair nanoribbon with two vacancies in a ``tail to tail'' configuration. The missing atoms
    belong to different sub-lattices and are indicated by the blue crosses. 
    The ribbons consist of $N_1 \times N_2$ cells with $N_1$ counting the cells in $x$-direction and $N_2$ in $y$-direction.
    In the lateral direction periodic boundary conditions are imposed at the edges.}
  \label{fig:ribbon}
\end{figure}
\begin{figure*}[t]
    \centering
    \includegraphics[height=0.33\textwidth]
             {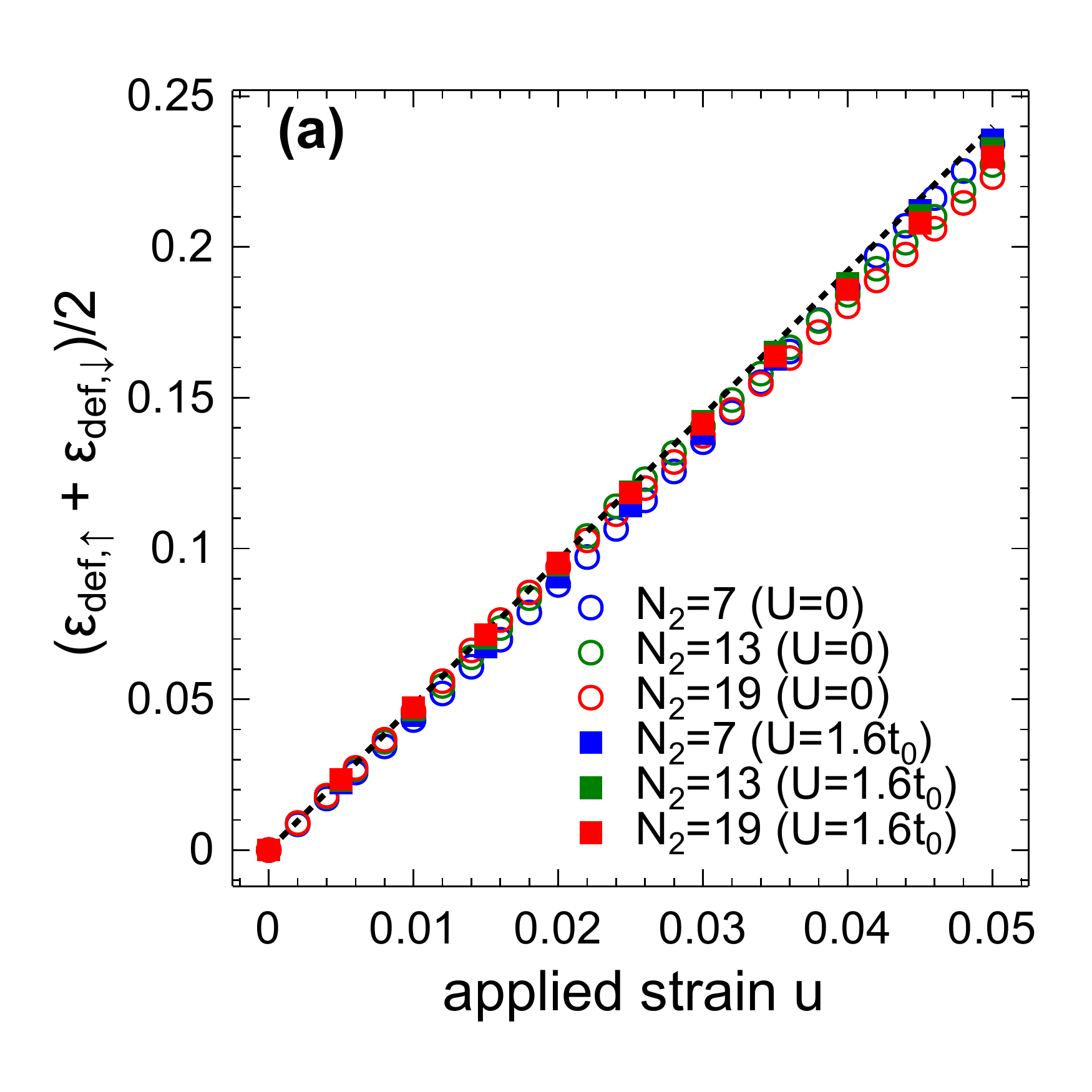}\qquad
    \includegraphics[height=0.33\textwidth]
             {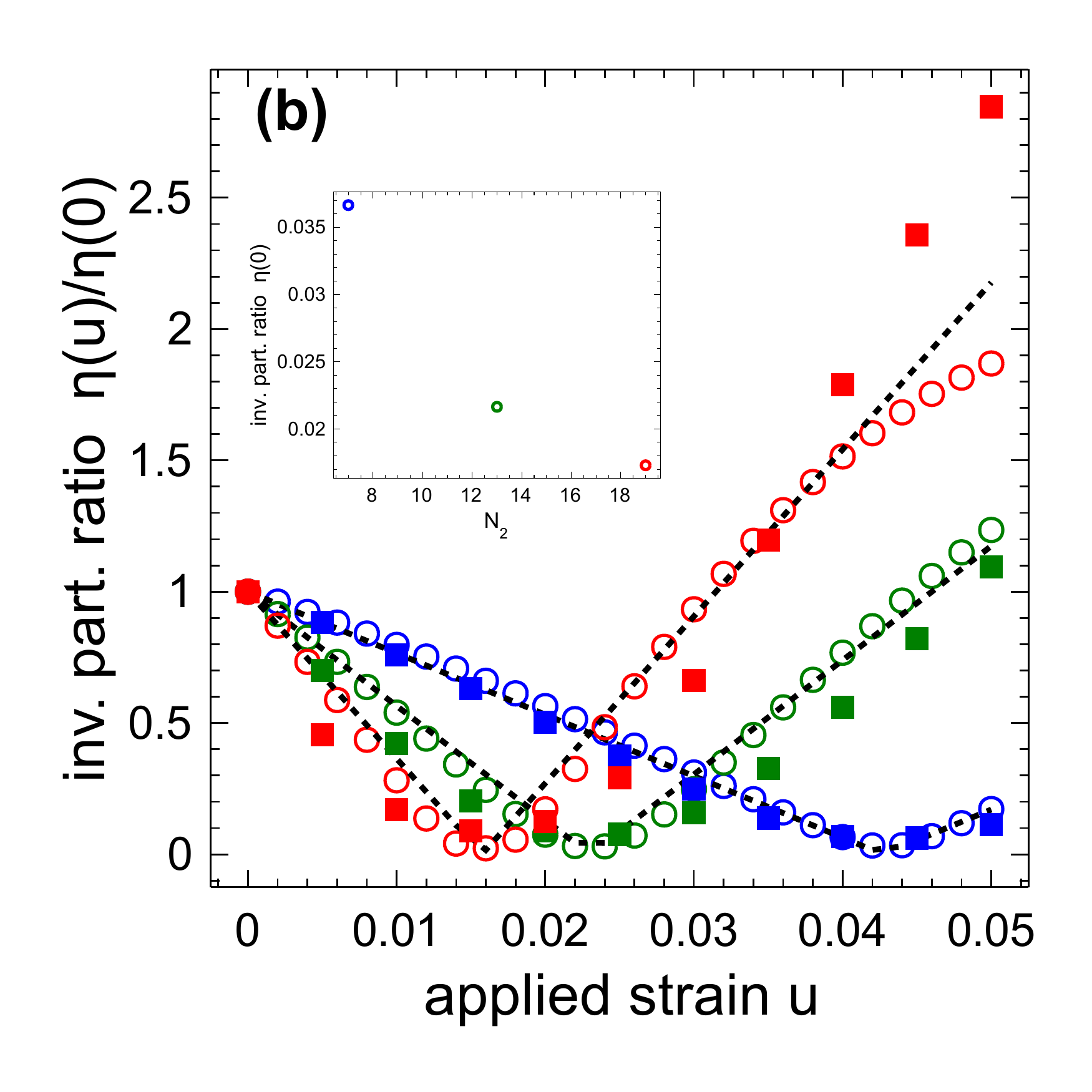}\\
    \includegraphics[height=0.33\textwidth]
             {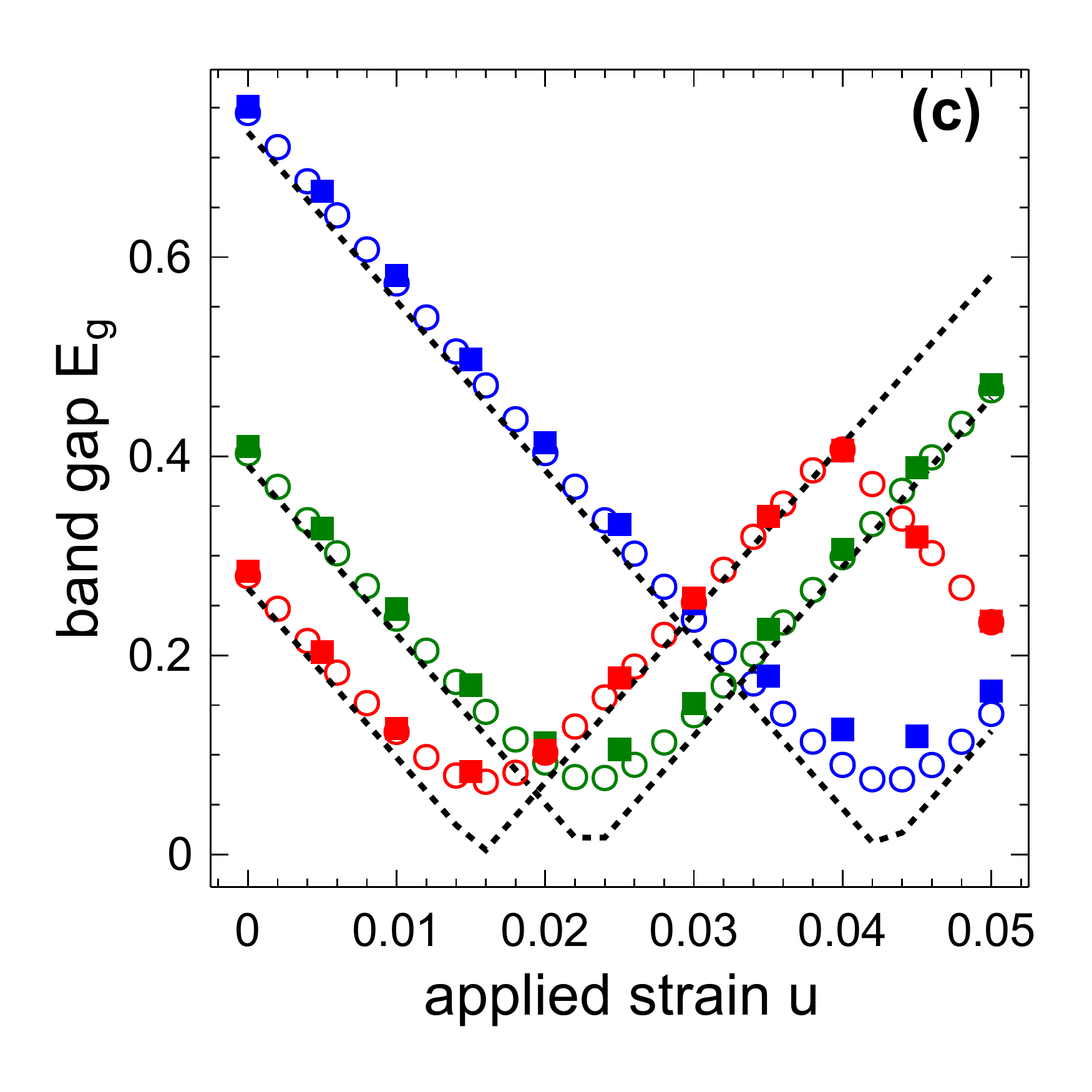}\qquad
    \includegraphics[height=0.33\textwidth]
             {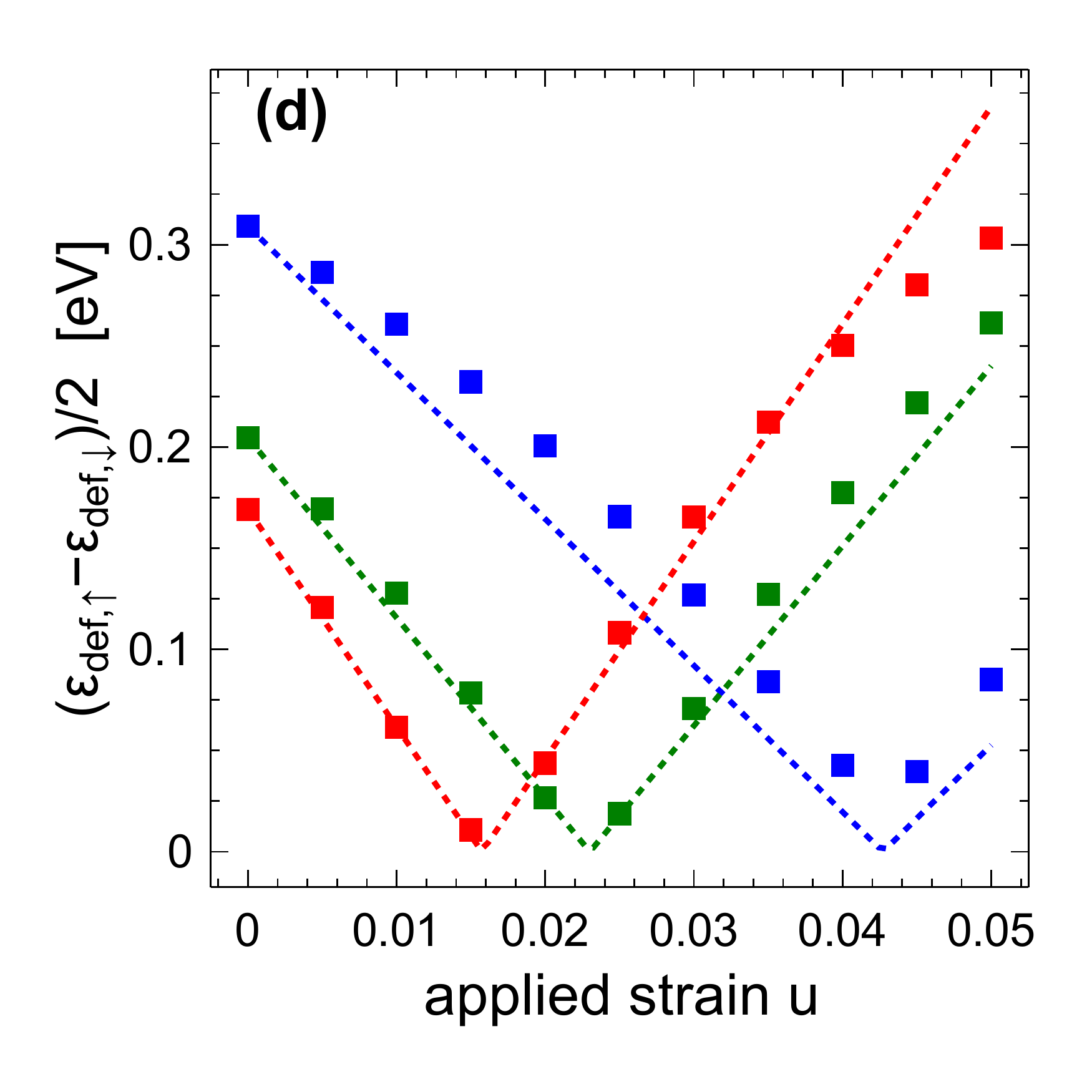}           
             \caption{Strain dependence of (a) the (average) defect energy $(\varepsilon_{\rm def.,\sup}+\varepsilon_{\rm def.,\sdown})/2$ 
                 of the localized state,
            (b) the normalized participation ratio $\eta(u)/\eta(0)$, (c) the band gap $E_{\rm g}$
            and (d) the exchange splitting $\varepsilon_{\rm def.,\sup}-\varepsilon_{\rm def.,\sdown}$
            for AGNRs of different widths. Inset in (b) shows the dependence of $\eta(0)$ on the ribbon width. Symbols denote results obtained from a TB calculation
            and dashed lines indicate behavior according to Eq.\ \eqref{eq:gap}. Here $N_1=100$ (for $U=0$),  $N_1=60$ (for $U=1.6t_0$) 
        and $N_2=7,13,19$.}
    \label{fig:single}
\end{figure*}
%%
%%%%%%%%%%%%%%%%%%%%%%%%%%%%%%%%%%%%%%%%%%%%%%%%%%%%%%%%%%%%%%%%%%%%%%%%%%%%%%%
%% MODEL
%%%%%%%%%%%%%%%%%%%%%%%%%%%%%%%%%%%%%%%%%%%%%%%%%%%%%%%%%%%%%%%%%%%%%%%%%%%%%%%
\section{Model}
We consider ribbons with armchair edges as sketched in Fig.\ \ref{fig:ribbon}.
In general, the defects can be either (single-atom) vacancies or chemisorpted hydrogen atoms\cite{yahe07}, which both effectively lead to
the removal of $p_z$ orbitals. In the following, ``vacancy'' will refer to both types of defects. We note that removing a
carbon atom also leads to an unpaired $\sigma$-electron. Experimental results\cite{nats+13} suggest that this
electron remains unbonded without hybridizing with the extra $\pi$-electron. Thus, both electrons provide independent
contributions to the total magnetization. As in previous studies\cite{pafe+08,ya10} we only consider the $\pi$-electrons in the following.

Uniform strain is applied in the direction parallel to the armchair edges. We include the Poisson effect (narrowing of the ribbon upon stretching), and write the elements of the strain tensor $\ten{u}=[\strain_{xx}, \strain_{xy}; \strain_{xy}, \strain_{yy}]$ as $\strain_{xx}=\strain$, $\strain_{yy}=-\nu \strain$ and $\strain_{xy}=0$, where $\nu$ is the Poisson ratio $\strain$ the applied strain. The bond-vectors $\vec{r}_{ik}$ connecting atom $i$ and atom $k$ are transformed as $\vec{r}_{ik}\to\left(\ten{I} + \ten{\strain}\right)\cdot\vec{r}_{ik}$ with $\ten{I}$ being the $2\times2$ identity matrix. For a realistic interatomic potential, this relation is slightly modified\cite{mile+15}, and the edges of the vacancies will be reconstructed\cite{yahe07}. These effects depend on the details of the interatomic potential, and are not included here.%%

To model the electronic properties of the nanoribbon we use a Hubbard model, for which the total Hamiltonian is
$H = H_0 + U \sum_i c^\dagger_{i\sup} c_{i\sup} c^\dagger_{i\sdown} c_{i\sdown}$. 
Here, $c^\dagger_{is}$ creates an electron at site $i$ with spin $s=\{\sup,\sdown\}$ and $U$ denotes
the interaction strength.
The non-interacting part, $H_0$, is given by the usual tight-binding model with nearest neighbor
hopping (see for instance \cite{ramo+13}). Expanding the hopping amplitude $t_0(r)$ and the electron-ion potential $v_0(r)$ up to first
order in small displacements of the atoms, yields the following Hamiltonian\cite{suan02,cagu+09,voka+10}
%%
%\begin{multline}
\begin{equation}
\fl    H_0 = \sum_{s=\{\sup,\sdown\}}\sum_{i=1}^{N_{\rm at}} 
        \left[ 
            \left(\frac{g}{r_0} \sum_{<k,i>}\frac{\vec{r}_{ik} \cdot \ten{\strain} \cdot \vec{r}_{ik}}{r_0} \right) c^\dagger_{i s} c_{i s} %\right.\\\left.
            - t_{0}(r_0) \sum_{<j,i>} \left(1 - \frac{\beta}{r_0}\frac{\vec{r}_{ij} \cdot \ten{\strain} \cdot \vec{r}_{ij}}{r_0} \right)c^\dagger_{i s} c_{s j} 
        \right] \;. \label{eq:TBHamiltonian}
\end{equation}
%\end{multline}
%%
The first sum runs over all $N_{\rm at}$ atoms and the other sums are restricted to nearest neighbors of the atom at site $i$.
The on-diagonal contribution is given by the deformation potential and its strength $g/r_0 = \partial_r v_{0}|_{r_0}$. The modification
of the hopping due to displacement of the atoms is determined by $\beta/r_0 = -{\partial_r t_{0}}/{t_{0}|_{r_0}}$. In the following we use $t_0=2.8\;{\rm eV}$, $U=1.6 t_0$ \cite{scro+13}, $g=4\;{\rm eV}$, $\beta=3.37$ \cite{peca+09} and $\nu=0.2$\cite{pete09}.

%%%%%%%%%%%%%%%%%%%%%%%%%%%%%%%%%%%%%%%%%%%%%%%%%%%%%%%%%%%%%%%%%%%%%%%%%%%%%%%
%% SINGLE VACANCY
%%%%%%%%%%%%%%%%%%%%%%%%%%%%%%%%%%%%%%%%%%%%%%%%%%%%%%%%%%%%%%%%%%%%%%%%%%%%%%%
\section{Results}
\subsection{Single vacancy.}
First we consider the case of a single vacancy at the center of the ribbon for $U=0$. Since the corresponding localized defect state
is always located at the Dirac point\cite{wa02}, we expect its energy to follow the behavior of 
the deformation potential, $v_{\rm D} = \frac{3 g}{2} \left( u_{xx} + u_{yy} \right)$. From Fig.\ \ref{fig:single}(a) we infer
that this is true for strains up to $\approx 3\%$. The localization of the defect state is quantified by the inverse participation
 ratio $\eta = \sum_i |\psi_i|^4$, where $\psi_i$ are the components of the normalized eigenfunction of the defect state. 
For a maximally localized state $\eta=1$, while $\eta\to0$ for completely
delocalized states and large systems. In Fig.\ \ref{fig:single}(b) the dependence of $\eta$ on the applied strain
and on ribbon width (inset) is shown. The localization at zero strain decreases with increasing width\cite{pafe+08}. 
For a given width, $\eta$ shows a non-monotonic strain-dependence, with a minimum that shifts toward higher strains for decreasing ribbon widths. This behavior can be understood in a qualitative way by recalling that the electronic energy in the Dirac picture is given by $E=\hbar v_{{\rm F}} k$, where $v_{{\rm F}}$ is the Fermi velocity and $k$ is the electron momentum which has the unit of inverse length\cite{cagu+09}. The relevant energy scale is set by the band-gap $E_{\rm g}$, and the corresponding length scale is consequently set by $\hbar v_{{\rm F}}/E_{\rm g}$. This suggests that the localization length scales inversely with the band-gap. For a semiconducting AGNR 
the band gap is approximately given by\cite{lugu10}
\begin{equation}\label{eq:gap}
    E_g\approx 3 t_0 \min_{n=0,1,\ldots}\left| \frac{\pi}{\sqrt{3} N_2} (n - 1/3) - \frac{\beta}{2}(1+\nu)\strain\right|\;,
\end{equation}
which yields the characteristic zig-zag behavior shown as dashed lines in Fig.\ \ref{fig:single}(c). Note that for the
chosen widths the gap is determined by $n=0$ (for $N_2=19$ the next sub-band starts to contribute around $\strain=0.04$).  
So indeed, comparing the behavior of the participation ratio with the dependence of the band gap on the strain, we find that $\eta(\strain)/\eta(0) \approx E_{\rm g}(\strain)/E_{\rm g}(0)$, shown as dashed lines in Fig.\ \ref{fig:single}(b).

In the interacting case, Lieb's theorem\cite{li89} predicts that the ground state has total spin $1/2$. The 
energies $\varepsilon_{\rm def., s}$ are, in general, no longer degenerate and the exchange splitting
$\Delta_{\rm ex.} = \varepsilon_{\rm def.,\sup}-\varepsilon_{\rm def.,\sdown}$ is finite. In the ground state only one of the spin states
is occupied, which results in a finite magnetization. We have verified this by performing mean field calculations for
the Hubbard Hamiltonian $H$. In Figs.\ \ref{fig:single}(a)-(c) we show results for the average defect energy 
$(\varepsilon_{\rm def.,\sup}+\varepsilon_{\rm def.,\sdown})/2$, the inverse participation ratio and the band gap. One can see that
the observations made for $U=0$ also hold in the interacting case. In particular, the relation $\eta(\strain)/\eta(0) \approx E_{\rm g}(\strain)/E_{\rm g}(0)$ is still valid. 
Figure \ref{fig:single}(d) displays the strain dependence of the exchange splitting. Up to a certain strain it monotonically decreases
and then increases again. Comparing with the band gap shows that also $\Delta_{\rm ex.}$ follows the strain dependence
of $E_{\rm g}$ given by Eq.\ \eqref{eq:gap}. 
Altogether, we have shown that strain can be used to modify the (localization) properties of the defect state and
to tune the exchange splitting.

\begin{figure*}[t]
    \centering
    \includegraphics[width=0.32\textwidth]
             {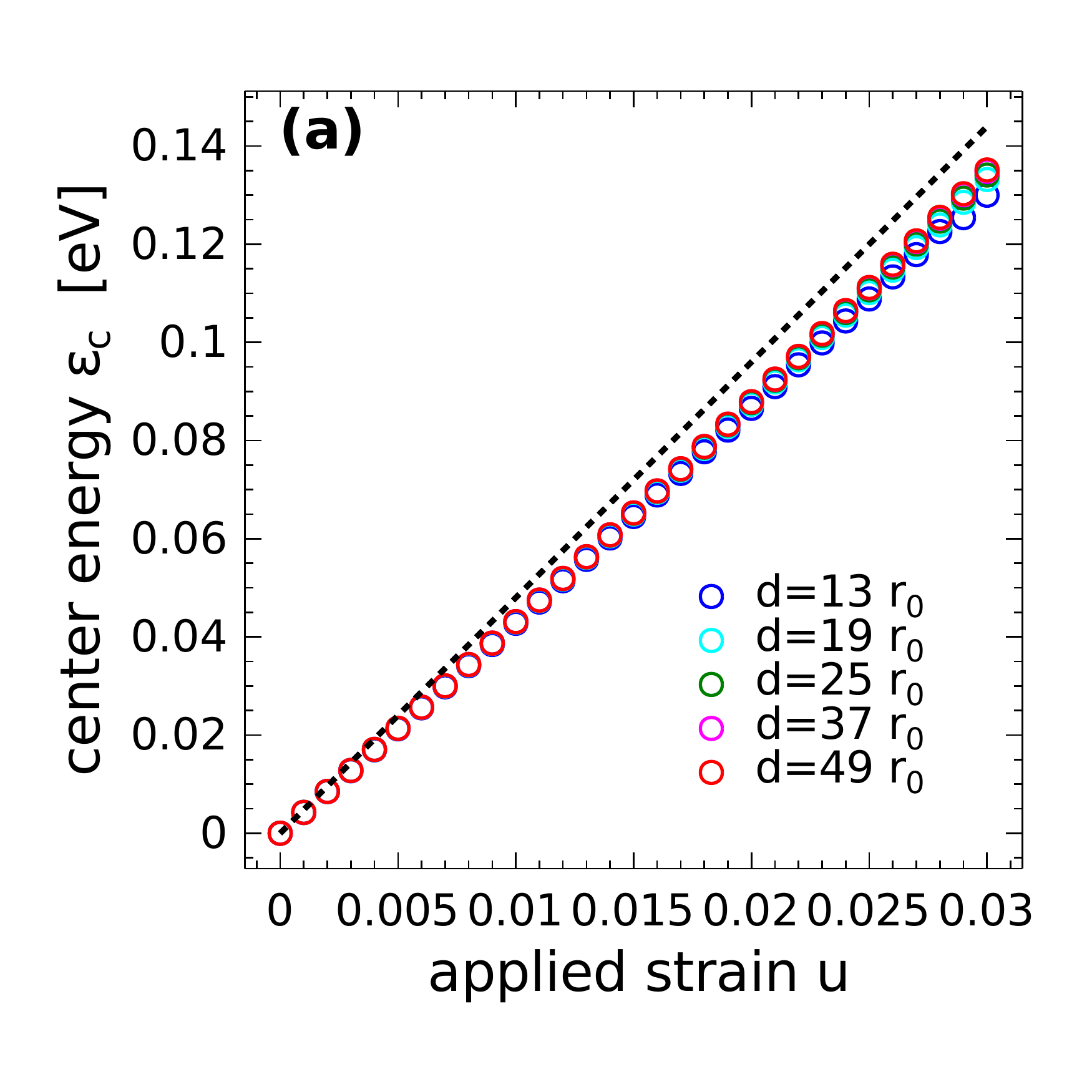}           
    \includegraphics[width=0.32\textwidth]
             {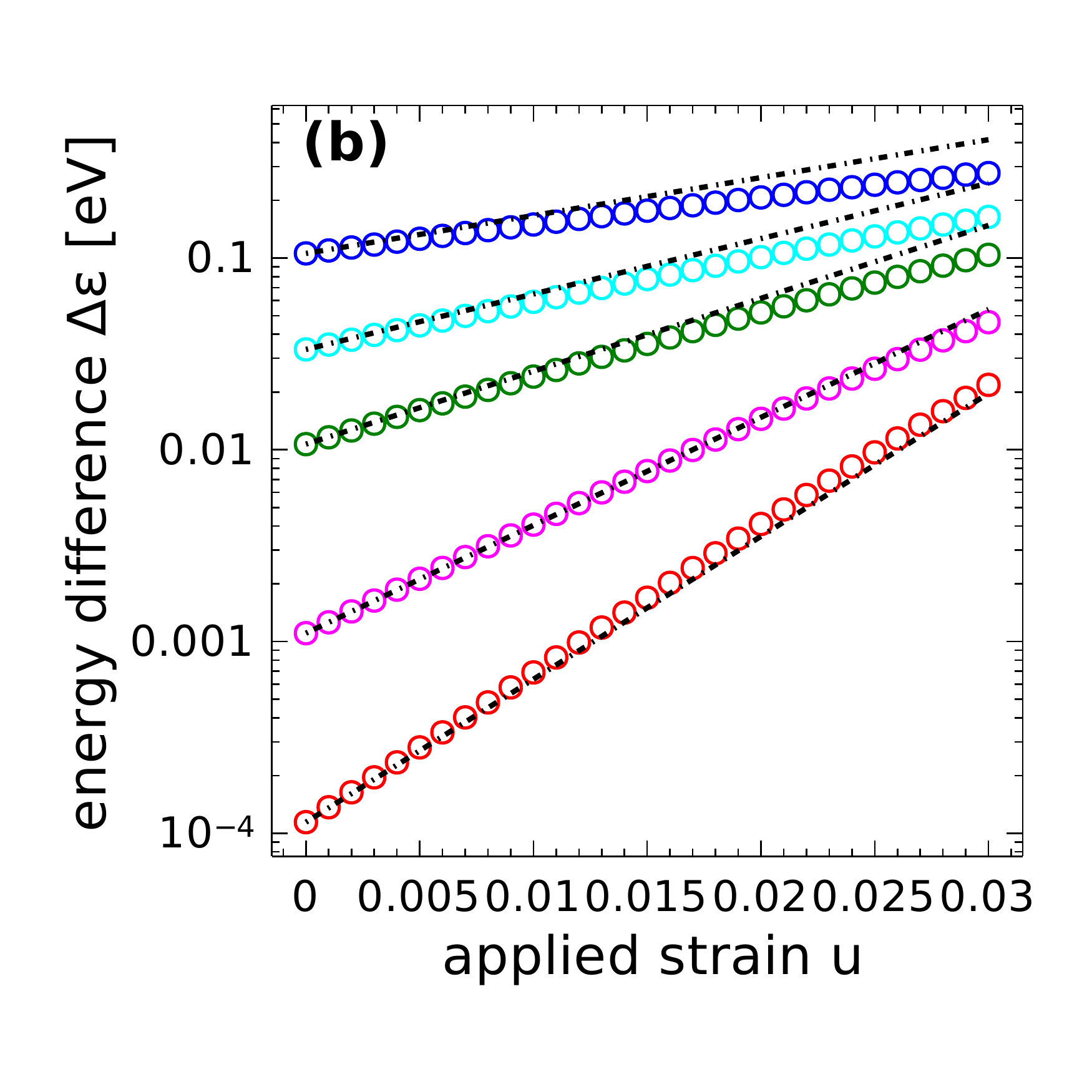}           
    \includegraphics[width=0.32\textwidth]
             {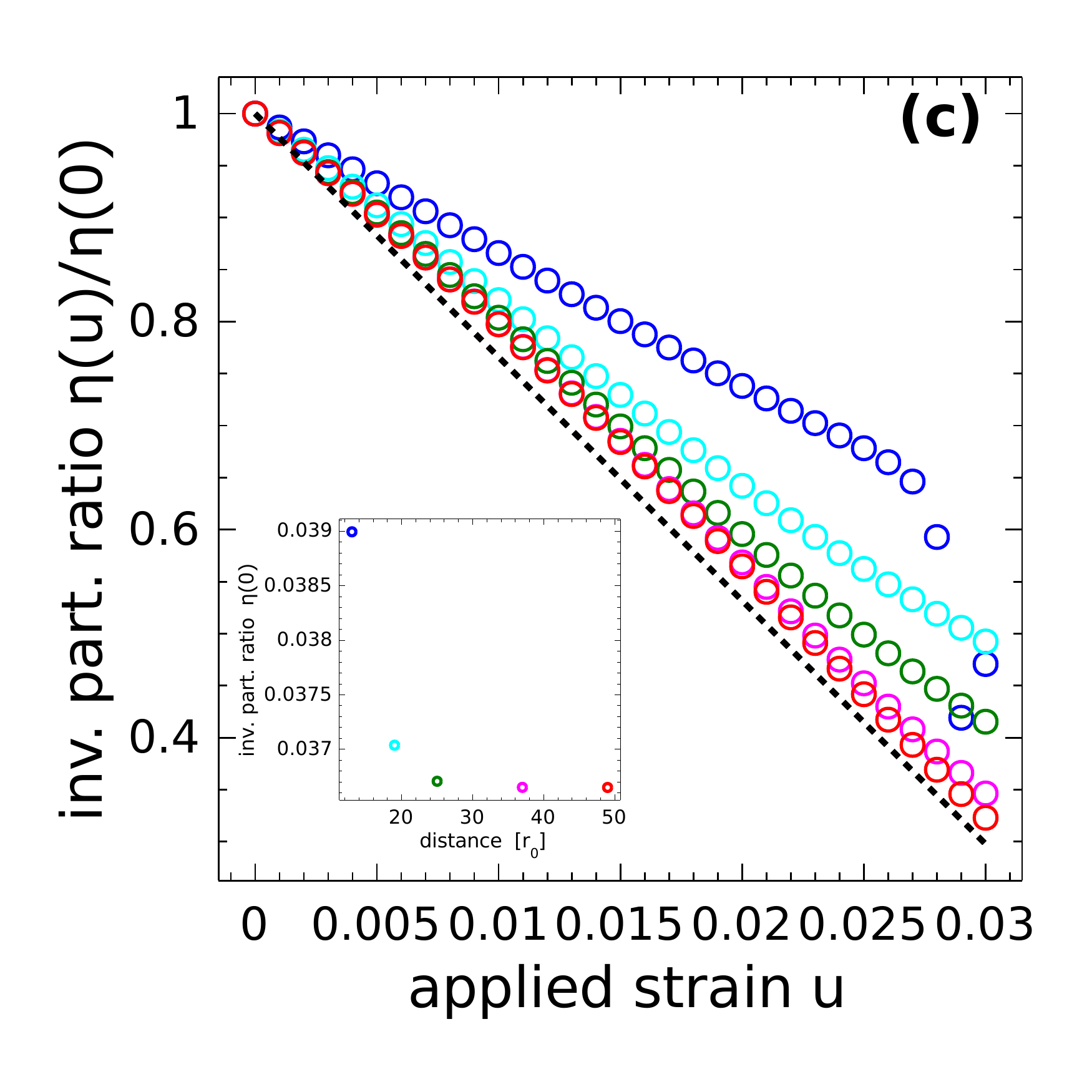}           
    \caption{Strain dependence of (a) the average energy $\varepsilon_{\rm c}$ and (b) the energy difference 
            of the two defect states, and
            (c) the normalized participation ratio $\eta(u)/\eta(0)$ for an AGNRs with two
            vacancies with distance $d$. Symbols denote results obtained from a TB calculation with $N_1=100$ and $N_2=7$.
            The dashed lines in (a) and (c) indicate behavior according to $v_{\rm D} = {3 g}/{2} (1-\nu)\strain$ and 
            $E_{\rm g}(\strain)/E_{\rm g}(0)$ with $E_{\rm g}$ from Eq.\ \eqref{eq:gap}, respectively. The dashed-dotted lines
            in (b) show $\Delta\varepsilon(\strain) = \Delta\varepsilon(0) \exp\left( \sqrt{3}\frac{\beta}{2r_0}(1+\nu)\strain d \right)$
            and the inset in (c) displays $\eta(0)$ vs the distance $d$.}
    \label{fig:double}
\end{figure*}
%%
%%%%%%%%%%%%%%%%%%%%%%%%%%%%%%%%%%%%%%%%%%%%%%%%%%%%%%%%%%%%%%%%%%%%%%%%%%%%%%%
%% DOUBLE VACANCY
%%%%%%%%%%%%%%%%%%%%%%%%%%%%%%%%%%%%%%%%%%%%%%%%%%%%%%%%%%%%%%%%%%%%%%%%%%%%%%%
\subsection{Two vacancies.}
Having characterized the influence of strain on the magnetic properties of a single vacancy, we now add a second vacancy such that their center of mass is at the center of the ribbon. This construct can be thought of as a simple realization of a spin qubit. We only consider vacancies with
missing atoms in different sub-lattices and in the ``tail to tail'' configuration (see Fig.\ \ref{fig:ribbon}). We first characterize the double-vacancy system in the non-interacting case, and use the obtained results to construct a double quantum dot model for the interacting system.

In the non-interacting case, we find two defect states
close to the band center\cite{pafe+08}. Linear combinations of the corresponding eigenstates yield states localized at the left and
right vacancy, respectively. Figs.\ \ref{fig:double}(a) and \ref{fig:double}(b) show the average energy of the two states and
their energy difference $\Delta\varepsilon$ for a ribbon with $N_2=7$. The average energy again approximately follows the behavior of the deformation 
potential and increases linearly with the applied strain. The energy difference at zero strain decreases exponentially with 
increasing distance of the vacancies\cite{pafe+08}. For a fixed width, the energy difference increases monotonically with
increasing strain. We find that it is approximately given by 
$\Delta\varepsilon(\strain) = \Delta\varepsilon(0) \exp\left( c \frac{\beta}{2r_0}(1+\nu)\strain d \right)$, where $c\approx \sqrt{3}$.
However, as we have seen in Fig.\ \ref{fig:single}(c) the band-gap decreases with strain and therefore at some
point the defect states start to overlap with the continuum. For the strains shown in Fig.\ \ref{fig:double}(b) this only 
happens for the smallest distance (shown as blue circles), since here the splitting is largest. The inverse participation ratio, shown in Fig.\ \ref{fig:double}(c), displays a pronounced
decrease when the defect states are shifted into the continuum (blue circles). Otherwise $\eta$ is found to depend only weakly on the distance (see inset of Fig.\ \ref{fig:double}(c))  
and it decreases approximately linearly with the applied strain, but with a distance dependent slope. For vacancies which are
far apart the behavior is well described by the ratio $E_{\rm g}(\strain)/E_{\rm g}(0)$, as expected for nearly independent vacancies.

\subsection{Double quantum dot model.}
According to Lieb's theorem\cite{li89} the ground state of the ribbon for $U>0$ with two single-atom vacancies in different sub-lattices
has total spin zero $S=0$ (singlet state)\cite{pafe+08}. In this case the highest occupied state, which corresponds to the lower lying defect state $\psi^{(-)}$,
is filled with two electrons of opposite spin. To estimate the energy gap to the triplet states ($S=1$), we only consider the defect 
states $\psi^{(\pm)}$ and take them as highest occupied and lowest unoccupied orbitals\cite{lihe+14,drbu15}. 
By forming linear combinations of $\psi^{(\pm)}$, 
we obtain the states $\psi^{(\rm L/R)}$ localized at the respective vacancy. Defining corresponding annihilation operators 
$a_{ns} = \sum_i \psi^{(n)}_i c_{is}$, where $n={\rm L,R}$, we can write the Hamiltonian for the double quantum dot model as

\begin{equation}
    \label{eq:DQDHam}
\fl    H_{\rm DQD} = \sum_s \left( \varepsilon_{\rm L} a_{{\rm L} s}^\dagger a_{{\rm L} s} +
                                \varepsilon_{\rm R} a_{{\rm R} s}^\dagger a_{{\rm R} s} %\right. \\ \left.
                                - t  \left( a_{{\rm L} s}^\dagger a_{{\rm R} s} +  a_{{\rm R} s}^\dagger a_{{\rm L} s} \right)
                 \right) + U_{\rm eff} \sum_n a_{n\sup}^\dagger a_{n\sup} a_{n\sdown}^\dagger a_{n\sdown}\;.
\end{equation}

Here, $\varepsilon_{\rm L}=\varepsilon_{\rm R}$ equals the average frequency of the two defect states, $t=\Delta\varepsilon/2$ quantifies
the tunnel coupling of the local states and $U_{\rm eff} = U \eta$ is the local interaction strength.

In the local basis $\ket{s_{\rm L}, s_{\rm R}}$, where $s_{\rm L/R}=\{\sup,\sdown\}$ denotes the 
spin state of the left and right vacancy, the singlet state of the two-state model
is given by $\Psi_{S=0} = (\ket{\sup,\sdown} - \ket{\sdown,\sup})/\sqrt{2}$ and the triplet states are
$\Psi^{(0)}_{S=1} = (\ket{\sup,\sdown} + \ket{\sdown,\sup})/\sqrt{2}$, $\Psi^{(1)}_{S=1} = \ket{\sup,\sup}$ and 
$\Psi^{(-1)}_{S=1} = \ket{\sdown,\sdown}$. The respective energies are found from Eq.\ \eqref{eq:DQDHam} to be
\begin{subequations}\label{eq:DQDEnergies}
\begin{eqnarray}
    E_{S=0} &={}& \frac{U_{\rm eff}-\sqrt{U_{\rm eff}^2 + 16 t^2}}{2} \equiv J,\\
    E^{(0)}_{S=1} &={}&E^{(1)}_{S=1}=E^{(-1)}_{S=1} = 0\;.
\end{eqnarray}
\end{subequations}
For $|t|\ll U_{\rm eff}$ the exchange energy can be approximated by $J\approx -4t^2/U_{\rm eff}$.

\begin{figure}[bt]
    \centering
    \includegraphics[width=0.5\columnwidth]
             {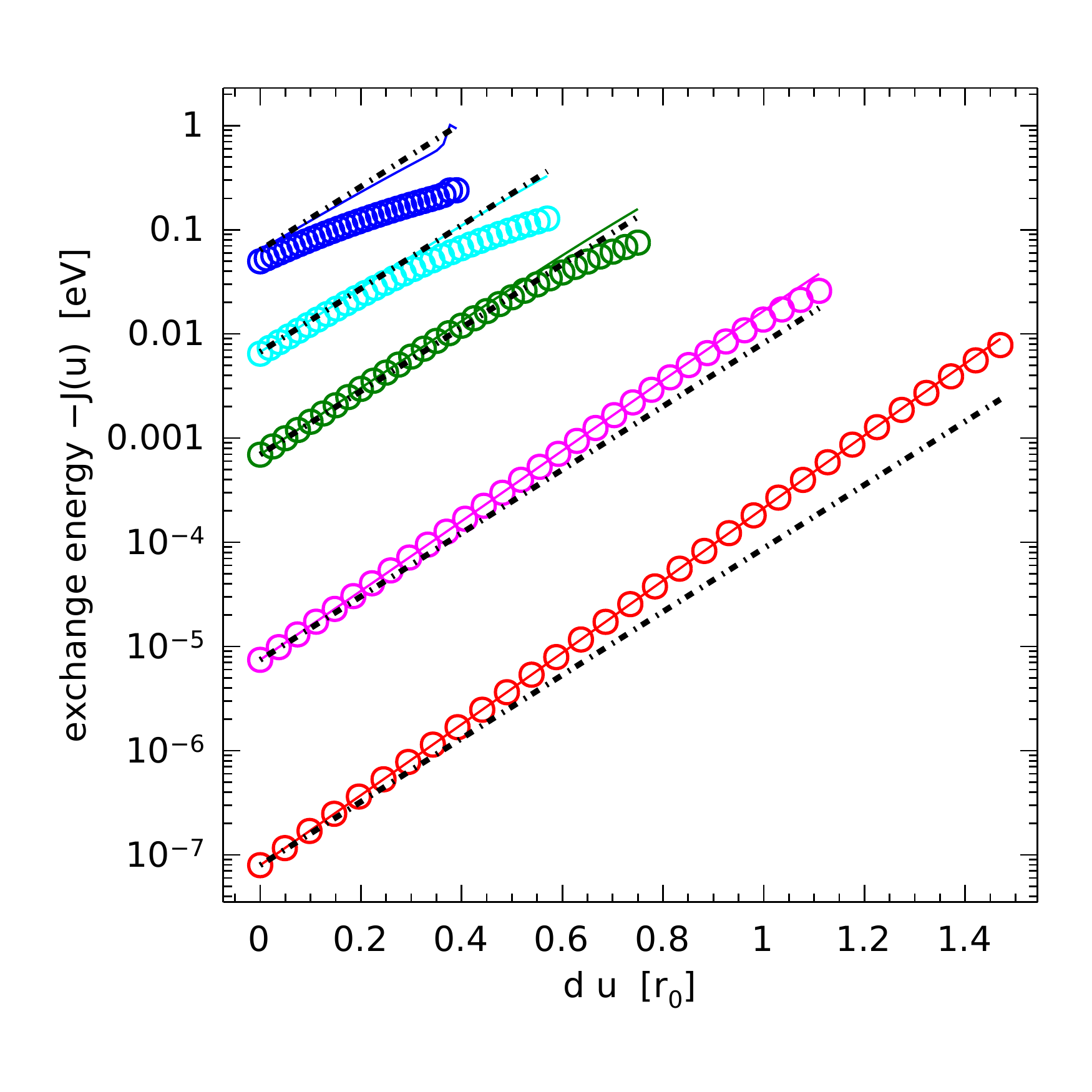}           
             \caption{Strain and distance dependence of the exchange interaction $J$ for an AGNR ($N_1=100$ and $N_2=7$) with two
                 vacancies with distance $d$. Symbols denote results obtained from the DQD model \eqref{eq:DQDEnergies}
                 and data shown in Fig.\ \eqref{fig:double}. Full lines show the strong interaction approximation
                 $J\approx -4t^2/U_{\rm eff}$ and dashed-dotted lines denote 
                 $J(\strain)=J(0) \exp\left( 2\sqrt{3}\frac{\beta}{2r_0}(1+\nu)\strain d \right)$.
             }
    \label{fig:exchange}
\end{figure}

Using the data shown in Fig.\ \ref{fig:double} we calculate the exchange energy $J$ as a function of the vacancy distance
and the applied strain. The resulting behavior is shown in Fig.\ \ref{fig:exchange}. Note that we plot $J$ vs $d\strain$. 
For zero strain $|J|$ decreases exponentially with increasing distance. 
Moreover, we find that $|J|$ is exponentially increasing with increasing $d\strain$, which is consistent with our
observation that $\Delta\varepsilon\propto\exp(2\kappa d\strain)$ with $\kappa\approx\sqrt{3}{\beta}(1+\nu)/2r_0$. 
Since the participation ratio depends only weakly on the
vacancy distance, but the energy difference strongly decreases, the behavior of $J$ approaches the strong interaction
limit $|t|\ll U_{\rm eff}$ for increasing distance. Thus the exchange interaction can be tuned over a wide range by adjusting 
the distance and/or by applying strain to the ribbon.
\section{Conclusions}
In summary, we have characterized the influence of geometry and strain on the electronic and magnetic properties
of semiconducting AGNRs with one and two vacancies. 
For a single vacancy we find that the degree of localization and the spin-exchange splitting follows the
strain-dependence of the band gap, which is non-monotonously changing. This simple relation connecting the intrinsic vacancy induced magnetism to strain via the band-gap has, to our knowledge, not been reported previously.
It implies that the spin-exchange can be changed to large extent by applying strain. Further, the singlet-triplet splitting
 in the two vacancy system depends exponentially on
the product of strain and inter-vacancy distance. This sensitivity with respect to strain may be exploited in the development of quantum-information and sensing applications. In an array of such vacancy-induced DQDs, the strain-tuning of the exchange interaction can in principle be used to gain individual control of the magnetic properties.
Combining strain-sensitive magnetic defects with mechanical resonators provides a viable route toward magneto-mechanical devices\cite{gakl+13} and might make it possible to couple vacancy-induced DQDs over long distances\cite{buim06}.

%%%%%%%%%%%%%%%%%%%%%%%%%%%%%%%%%%%%%%%%%%%%%%%%%%%%%%%%%%%%%%%%%%%%%%%%%%%%%%%
%%% BIBLIOGRAPHY
%%%%%%%%%%%%%%%%%%%%%%%%%%%%%%%%%%%%%%%%%%%%%%%%%%%%%%%%%%%%%%%%%%%%%%%%%%%%%%%
\section*{References}
\bibliographystyle{unsrt}
\bibliography{magvacresub}

\end{document}